\documentclass[final]{article}

\usepackage{neurips_2026}

\usepackage[utf8]{inputenc} 
\usepackage[T1]{fontenc}    
\usepackage{hyperref}       
\usepackage{url}            
\usepackage{booktabs}       
\usepackage{amsmath}        
\usepackage{amsfonts}       
\usepackage{amssymb}        
\usepackage{nicefrac}       
\usepackage{microtype}      
\usepackage{xcolor}         
\usepackage{algorithm}
\usepackage{algpseudocode}
\usepackage{graphicx}
\usepackage{subcaption}
\title{Exploring Bottom-Up Clustering for Creating Semantic IDs}

\author{%
  Leah Woldemariam 
  \\
  Department of Electrical and Computer Engineering\\
  Cornell Tech, Cornell University\\
  New York, NY, USA \\
  \texttt{lsw85@cornell.edu} \\
  \And
  Sudhanshu Garg \\
  PayPal \\
  San Jose, California, USA\\
  \texttt{sudgarg@paypal.com} \\
  \AND
  Taha Belkhouja \\
  PayPal \\
  San Jose, California, USA\\
  \texttt{tbelkhouja@paypal.com} \\
  \And
  Charles Kim-Yip \\
  PayPal \\
  San Jose, California, USA\\
  \texttt{cyip@paypal.com} \\
  \And
  Ali Sahami \\
  PayPal \\
  San Jose, California, USA\\
  \texttt{asahamishirazi@paypal.com} \\
}

\begin{document}

\maketitle

\begin{abstract}
    The success of generative retrieval has largely been attributed to the use of Semantic IDs, which improve over arbitrary item-level identifiers such as hashes by capturing the semantics of items.
    The main challenges faced when constructing Semantic IDs, however, is in mapping each identifier to a unique product and capturing information valuable to downstream tasks. Past works have appended additional codewords to de-duplicate item identifiers and utilized residual quantization to create hierarchical clusters.
    In this work, we present an algorithm for generating Semantic IDs that ensure the identifiers are both unique and preserve the structure of the original embedding. Key to our work is the use of bottom-up clustering to preserve local structure in the embedding space, improving the clustering quality of the resulting Semantic IDs and their utility for downstream generative retrieval.
\end{abstract}

\vspace{-1em}
\section{Introduction}
Generative retrieval has been established as a method for creating recommendation systems that provides better generalization performance on new items \cite{rajput2023recommender}. Instead of matching user and item embeddings, generative retrieval takes user information as input and generates potential items by treating retrieval as a sequential task. User information might include previous item interactions or text queries for the task of search. Its main advantage includes avoiding the error propagation due to multi-stage filtering, as well avoiding popularity bias and cold-start issues \cite{lin2025recommendation, rajput2023recommender}. Since items in these sequential models must be quantized, a recent line of research has been directed towards approaches to creating these quantized identifiers. Using item semantics to create these identifiers has shown to be the main facilitator in providing better item generalization \cite{rajput2023recommender}.

Identifiers that encode item semantics, called Semantic IDs (SIDs), are usually a short sequence of integers built in a hierarchical fashion so that top-level codewords represent coarse, categorical information, and lower level codewords increase precision in identifying an item. The item hierarchy allows for a smaller search space at each decoding step \cite{tay2022transformer} and allows for a clear method for recommending diverse items \cite{rajput2023recommender}. Among methods used for generating Semantic IDs, Residual-Quantized Variational Autoencoders (RQ-VAEs) and Residual K-Means (R-KMeans) stand out as the most popular due to the use of residual quantization for enforcing the item hierarchy. %

The primary issues surrounding the creation of Semantic IDs include codebook collapse and item collision. Codebook collapse occurs when only a few codewords within a codebook at any given level in a hierarchy contain the vast majority of items, under-utilizing the entire codebook space, while collisions occur when a single SID maps to multiple items \cite{lin2025unified}. Together, these issues can prevent semantically distinct items from being assigned unique and informative identifiers. Arbitrary assignment methods such as hashing or plain enumeration satisfy the property of uniqueness, but they fail to encode semantic relationships between items. Generating good SIDs therefore requires balancing two objectives: assigning each item a unique identifier while ensuring that semantically similar items are mapped to nearby Semantic IDs.

In this work, we propose an approach to creating Semantic IDs that both satisfies uniqueness and preserves distance in the embedding space. Using bottom-up rather than top-down clustering, we assign items to leaf nodes first, this ensures that each item corresponds to a single leaf node (uniqueness) and similar items in the embeddings space are inherently assigned nearby SIDs. The motivation behind the use of a bottom-up clustering is to preserve this local structure throughout the SID hierarchy: two items with similar embeddings will more likely share the same $(L-1)$-th codeword when there are $L$ codewords, thereby preserving fine-grained semantic relationships that can be exploited by downstream models. In contrast to conventional methods, which often introduce uniqueness through an ad hoc deduplication step and prioritize global structure during quantization, our approach preserves local neighborhoods by deisgn. By jointly preserving uniqueness and local embedding structure, our approach provides a more faithful discrete interface between pretrained item representations and downstream retrieval decisions, directly addressing the broader question of how pretrained representations can be made more actionable for downstream tasks.


\vspace{-1em}
\section{Related Work}\label{relatedwork}

\vspace{-1em}
\noindent \textbf{Semantic IDs.}
Most work on SIDs use Residual Quantization, with an earlier model using Vector Quantization alone is one exception \cite{van2017neural}. \cite{ju2025generative} find that RK-Means can out-perform RQ-VAE, noting the added complexity does not necessarily lead to a significant increase in performance. Finally, \cite{liu2025understanding} finds that the structure of SIDs themselves may limit the scaling of GR models, and that combining SID embeddings with collaborative filtering does not improve scaling. \cite{xu2026mmq} introduces orthogonal regularization for multi-modal embeddings in order to ensure the weights learned by each expert do not learn redundant information.





\noindent \textbf{Collision and Codebook Utilization.}
Many works cite that the primary issue in creating SIDS through the RQ-VAE is the high collision rate and low codebook utilization \cite{kuai2024breaking}. \cite{kuai2024breaking} finds that not only can codebooks heavily concentrate products into a small number of codes, but that this phenomenon may be isolated into specific levels. \cite{lin2025unified} similarly note that one pitfall of the Semantic IDs is the lack of uniqueness of the Semantic IDs, which they solve by concatenating the Semantic ID with unique identifiers from the item's embedding. At Snap, \cite{ju2026semantic} finds the codebook collapse and collision are primary issues in creating SIDs, and mitigate these issues by backpropagating through the codebook and by introducing heuristics, respectively. Alternatively, semantics is usually measured and improved by categorical/tag information \cite{spotify}. \cite{liu2026cat} adds terms to the loss function that maximize dispersion and improves codebook utilization efficiency.
\vspace{-1em}

\section{Methods}\label{methods}

\vspace{-1em}

\noindent \textbf{Overview.} 
As in Algorithm \ref{alg:bottomsup} presented in the Appendix, the set of items set can be represented as an $N \times d$ matrix $X$. Each $x_i \in \mathbb{R}^d$ represents a number of text item attributes used as input together into a pre-trained transformer to create an embedding.
Denote the set of labels for each item $i$ at level $\ell$ as $g^{(\ell)}$, and $(g_i^{(1)}, \dots, g_i^{(L)})$ represents the Semantic ID for item $i$.

Unlike arbitrary IDs, learned SIDs often struggle to guarantee uniqueness. At the same time, products that are similar to each other should be assigned nearby SIDs in the identifier hierarchy. This is one of the key properties a SID should have and is critical for mitigating the cold-start problem, as it allows unseen or sparsely observed items to share identifier structure with semantically related products. Thus, mapping these embeddings to discrete identifiers can be viewed as a constrained hierarchical clustering problem: the resulting identifiers should uniquely distinguish individual items while preserving the local structure of the original embedding space.

We seek to hierarchically assign items their identifiers through bottom-up clustering, i.e., by first clustering items and successively merging them at each level in the hierarchy. By using bottom-up clustering, we preserve local structure within the embedding space and its fine-grained item relationships. In particular, nearby items are grouped together at the lowest levels of the hierarchy before these local groups are progressively merged into coarser clusters. This is in contrast to top-down approaches, which first capture global information in the embedding space and subsequently partition it into increasingly fine-grained regions, potentially losing local relationships introduced at deeper levels of the hierarchy. By preserving these local relationships, our approach aims to retain the information in pretrained item representations that is most useful for downstream retrieval decisions, providing a more actionable interface between latent representations and generative models.


\begin{figure}[h]
    \begin{subfigure}[c]{0.5\linewidth}
    \centering
    \includegraphics[width=1\linewidth]{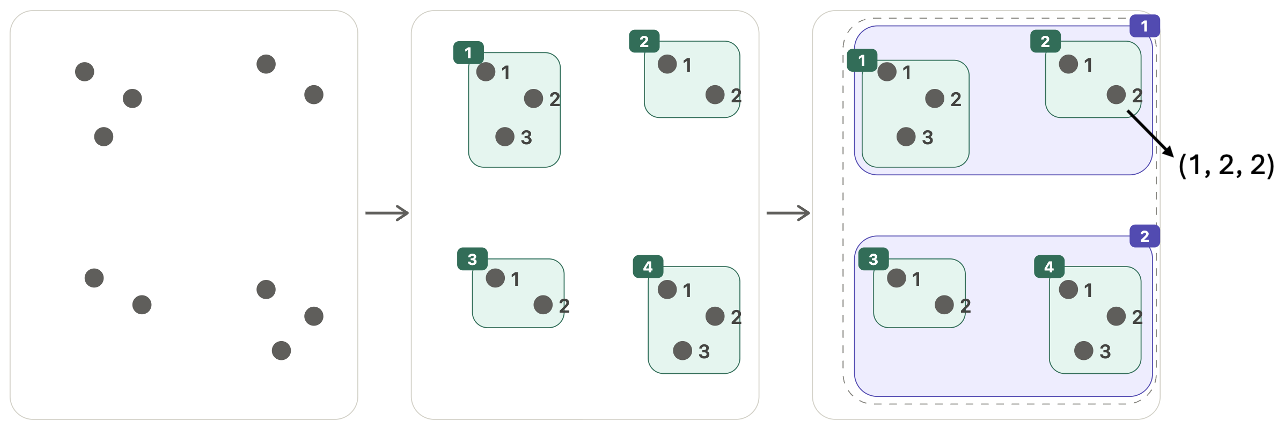}
    \end{subfigure}
    \hfill
    \begin{subfigure}[c]{0.5\linewidth}
    \centering
    \includegraphics[width=0.55\linewidth]{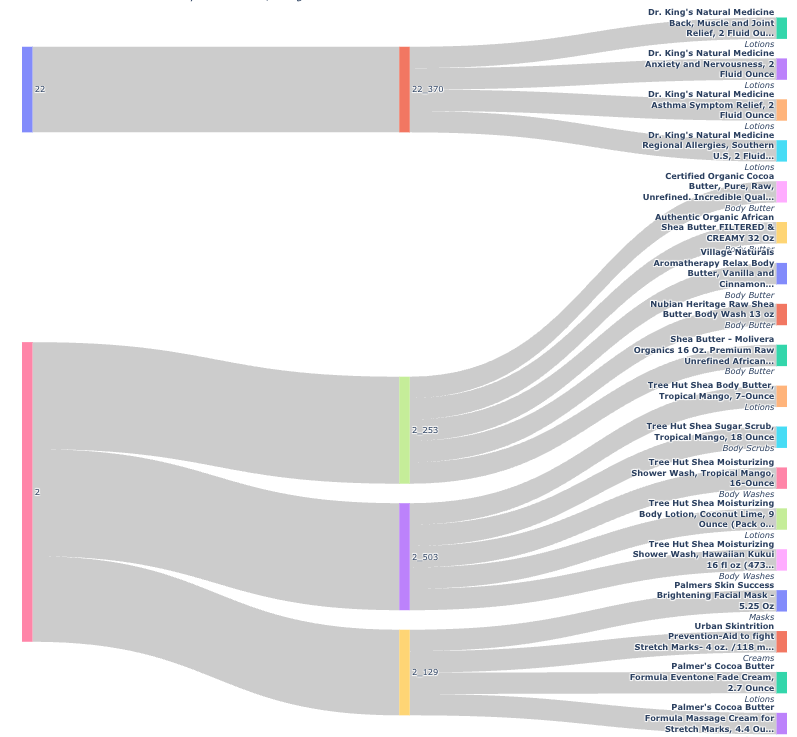}
    \end{subfigure}
    \caption{(a) A diagram depicting bottom-up clustering, with each node representing the embedding of an item. In the leftmost panel are the unclustered items, and in the middle panel is the initial clusters (green). The rightmost panel shows the final hierarchical clustering with the top-level groups in purple. (b) Example of a sample items assigned SIDs using the bottom-up approach.}
\end{figure}
\noindent \textbf{Algorithm.}
We construct a hierarchical, bottom-up clustering of the product embeddings $X \in \mathbb{R}^{N \times d}$ as described in Algorithm \ref{alg:bottomsup}. The procedure begins at the finest level by clustering the raw embeddings, streamed in chunks, to partition $N$ products into $k_0$ fine-grained groups. To ensure that the final codebook uniquely identifies each item, groups containing more than $K$ items are further split until every sub-cluster contains at most $K$ items. We enumerate each item within these fine clusters and assign its index as the $L$-th codeword, ensuring every product has a unique within-cluster identifier.


For each of the initial groups, we compute a weighted centroid using its member count, yielding a compact set of $k_1$ representative points that summarizes the full dataset. We then coarsen the hierarchy upward by iteratively merging these centroids. At each level, we cluster the current set of
centroids using standard Agglomerative Clustering, using each centroid's member count as a sample weight so that the merge respects the true population mass. After each merge, we recompute the centroids and weights for the newly formed groups, and propagate the new labels back down to every original product. Repeating this produces a sequence of increasingly coarse cluster assignments.  

Each cluster at every level is enumerated $1, \dots, k_{\ell}$ as the standard format of SIDs. Note that $k_{\ell}$ need not be the same at each level. Merging clusters at $g^{(\ell)}$ creates the $\ell-1$ codewords, thereby creating clusters by moving up the tree. See Algorithm \ref{alg:bottomsup} for a more detailed description.

New items are assigned SIDs according to their nearest neighbors. A new item inherits the first $L-1$ codewords of its nearest neighbor, with a unique $L$-th codeword assigned within the selected cluster. When a cluster exceeds a predefined threshold, it can be further split into sub-clusters. Cluster assignment may also use multiple nearest neighbors for greater robustness. Importantly, this procedure allows the SID hierarchy learned from pretrained item representations to generalize to previously unseen items without reconstructing the identifier space. By transferring local structure to new items, the resulting SIDs make pretrained representations directly actionable for downstream retrieval under cold-start settings.
\vspace{-1em}


\section{Experiments} \label{expts}

\vspace{-1em}
\noindent \textbf{Metrics.} We evaluate our approach with a number of metrics to measure codebook quality and clustering quality, and additionally evaluate our method over a next-product prediction task to assess the downstream utility of the resulting Semantic IDs. To measure codebok quality, we measure the average number of items assigned to each SID, which captures the extent to which multiple items share the same identifier and provides an indication of identifier collisions and codebook utilization. To measure clustering quality, we evaluate silhouette score and cosine similarity; we measure these metrics at each depth of the tree created by the SIDs, i.e. for clusters created by the first codebook, second codebook, and so on, while emphasizing the values at the final clusters. These metrics characterize how well the SID hierarchy preserves the structure of the pretrained embedding space.

When a dataset has categorical labels for each item, we additionally measure the average number of categories among items within a codeword, where fewer categories indicate greater semantic coherence within the resulting clusters. As in \cite{rajput2023recommender}, recall and NDCG is reported for item prediction tasks. These downstream metrics allow us to evaluate whether improvements in the structure of the Semantic IDs translate into better retrieval decisions, connecting representation quality to downstream actionability.

\noindent \textbf{Baselines.} We compare against an RQ-VAE, the a widely used method for constructing SIDs, as our primary baseline. The RQ-VAE constructs codebooks in a top-down, hierarchical fashion starting with a fixed codebook size $L$. 

\noindent \textbf{Datasets.} We compare our algorithm to the above baselines with the Amazon Product Reviews dataset, a collection of product reviews for items \cite{ni2019justifying}. We evaluate over the Beauty and Sports \& Outdoors splits. We additionally compare our algorithm gainst the same baselines on a custom dataset containing approximately 6 million items, each with titles, model-generated captions of corresponding images, categorical information, and brand information. For each product, we concatenate the available textual information and encode it using a Qwen model to obtain the pretrained item embedding from which its SID is constructed.

\begin{table*}[t]
  \centering
  \makebox[\textwidth][c]{%
  \begin{tabular}{lcccccc}
    \toprule
    & \multicolumn{2}{c}{\textbf{Custom Dataset}} & \multicolumn{2}{c}{\textbf{Beauty}} &
    \multicolumn{2}{c}{\textbf{Sports \& Outdoors}} \\
    \cmidrule(lr){2-3} \cmidrule(lr){4-5} \cmidrule(lr){6-7}
    \textbf{Metric} & \small RQ-VAE  & \small Bottom-Up & \small RQ-VAE  & \small  Bottom-Up &
    \small RQ-VAE  & \small Bottom-Up \\
    \midrule
    \multicolumn{5}{l}{\textbf{Codebook Quality}} \\
    \quad Avg. Rows per ID & 1.63 &  1 & 1.35  & 1 &
    1.49 & 1\\
    \midrule
    \multicolumn{5}{l}{\textbf{Clustering Quality}} \\
    \quad Avg. Silhouette Score & 0.00 &  0.29 & $-$0.07   & 0.07
    & 0.02 & 0.07 \\
    \quad Cosine Similarity     & 0.80 & 0.82 & 0.76  & 0.80
    & 0.88  & 0.79 \\
    \midrule
    \multicolumn{5}{l}{\textbf{Next Product Prediction}} \\
    \quad Recall@10 &  0.020  &  0.036   & 0.0509 & 0.0567 & 0.0328  & 0.0327 \\
    \quad NDCG@10 &  0.001   & 0.002    & 0.0255 & 0.0296 & 0.0169   &  0.0166\\
    \midrule
    \multicolumn{5}{l}{\textbf{Other}} \\
    \quad Avg. Unique Categories & 1.06  & 14.30 & 1.16  & 3.75 & 1.21  & 4.54 \\
    \bottomrule
  \end{tabular}}
  \caption{Comparison of RQ-VAE and Bottom-Up clustering methods across the 5.8M Item dataset and Amazon Beauty datasets. Cluster-based metrics are reported at the final depth of the tree.}
  \label{tab:depthL}
  
\end{table*}

\noindent \textbf{RQ-VAE \& Next Product Prediction Model Implementation.} We first perform a hyper-parameter search on the batch size, learning rate, weight decay, codebook mode, codebook size, number of codebooks, and VAE hidden dimensions to compare our approach against the best possible RQ-VAE. The RQ-VAE used in these experiments has a codebook size of 256 for 3 codebooks. An extra codeword was appended to the RQ-VAE generated SIDs so that each item corresponding to a single SID, following the style of \cite{rajput2023recommender, ju2025generative}.

\noindent \textbf{Results.}
Table \ref{tab:depthL} reports the metrics for the finest-grained groups. For the bottom-up approach, this is evaluated at level $L-1$, corresponding to clusters defined by the tuples $(g^{(1)}, \dots, g^{(L-1)})$, since the $L$-th level always maps to unique items.

The bottom-up approach achieves higher silhouette scores across all datasets at the finest granularity, indicating that its finest-level clusters are more compact and better separated in the pretrained embedding space. The gap is particularly large on the custom dataset, which contains substantially more items and greater product diversity than the Amazon Reviews Beauty dataset. This suggests that preserving local structure may become increasingly important as the item space grows more heterogeneous. The bottom-up approach also achieves higher Recall@10 and NDCG@10 on the downstream next-item prediction task, showing that improved SID structure translates to better downstream retrieval. The bottom-up approach also yields a higher average number of unique categories within each cluster, which is expected because the optimal RQ-VAE configuration uses a codebook size substantially larger than the number of clusters at each level in the bottom-up hierarchy.

Unlike RQ-VAE, which learns a residual hierarchy from coarse to fine, our bottom-up method constructs the SID hierarchy directly from local relationships in the pretrained embedding space. We show that this structure-preserving approach leads to better clustering quality as well as improved performance on the downstream task of next-item prediction. More broadly, these results suggest that the way pretrained representations are converted into discrete abstractions materially affects their downstream utility: preserving tak-relevant local structure can make those representations more actionable for downstream decision-making.

\pagebreak
\bibliographystyle{alpha}
\bibliography{ref}





\pagebreak
\appendix
\section{Appendix}

Below is the algorithm describing our bottom-up approach. In our experiments, the bottom-up clustering approach used 2000 and 64 clusters at level $L-1$ and $L-2$, respectively for the custom dataset and otherwise matches the codebook size used by the RQ-VAE for the Amazon datasets.

\begin{algorithm*}[h]
\caption{Bottom-Up Hierarchical Clustering}\label{alg:bottomsup}
\begin{algorithmic}[1]
\Require Embedding matrix $X \in \mathbb{R}^{N \times d}$;
         strictly decreasing coarsening schedule
         $K = [k_1, k_2, \dots, k_L]$;
         optional per-cluster size cap $c_{\max}$.
\Ensure  Per-item level labels $\{g^{(0)}, g^{(1)}, \dots, g^{(L)}\}$, each
         $g^{(\ell)} \in \{0,1,\dots\}^{N}$ giving every item $i$ a label at level
         $\ell$; tuple $(g^{(1)}_i, g^{(2)}_i, \dots, g^{(L)}_i)$ is unique per item.
\Statex
\Statex \textbf{Step 1: Finest clustering}
\State $g^{(L-1)} \gets \textsc{MiniBatchKMeans}(X,\, k_1)$
       \Comment{returns a label $g^{(L-1)}_i \in \{0,\dots,k_1-1\}$ for every item $i$}
\State $(C, w) \gets \textsc{GroupCentroids}(X,\, g^{(L-1)})$
       \Comment{$C \in \mathbb{R}^{k_{\ell} \times d}$ centroids; $w$: its item count}
        \Comment{$a$ maps cluster labels at $\ell+1$ to labels at level $\ell$}

\ForAll{groups $s$ in $g^{(L-1)}$}
    \State assign the members of $s$ distinct ids $0,1,2,\dots$ (running count within $s$) as their label $g^{(L)}$
\EndFor
\State $g^{(L)} \gets$ these per-item ids
       \Comment{(uniqueness)}

\Statex \textbf{Step 2: Recursive Merging}
\For{$\ell = L-2$ \textbf{to} $1$}
    \State $g^{(\ell)} \gets \textsc{MiniBatchKMeans}\big(C,\, k_\ell,\,
            \text{sample\_weight}=w,\ \text{parent}=g^{(\ell+1)}\big)$
           \Statex \hspace{\algorithmicindent}
    \If{$c_{\max} \neq \varnothing$}
         $g^{(\ell)} \gets \textsc{EnforceMaxClusterSize}\big(X,\, g^{(\ell)},\, c_{\max}\big)$
               \hspace{\algorithmicindent}\Comment{ splits any level-$\ell$
               group with more than $c_{\max}$ items}
    \EndIf
    \State relabel $g^{(\ell)}$ to dense ids $\{0,\dots,n_\ell-1\}$
    \State $(C, w) \gets \textsc{GroupCentroids}(X,\, g^{(\ell)})$
\EndFor
\State \textbf{return} $(g^{(1)}, g^{(1)}, \dots, g^{(L)})$
\end{algorithmic}
\end{algorithm*}



\end{document}